\documentclass[twocolumn,10pt,letterpaper]{article}
\pdfoutput=0
\usepackage{amsopn}
\usepackage{usenix}
\usepackage{graphicx}
\usepackage{caption}
\usepackage[binary,squaren]{SIunits}
\usepackage{tikz}
\usepackage[hyphens]{url}
\usepackage{times}
\usepackage{tabularx}
\usepackage[sort,numbers]{natbib}
\usepackage{enumitem}
\usepackage{amssymb}
\usepackage{listings}
\usepackage[font={small}]{caption}
\usepackage{bbding}
\usepackage[yyyymmdd,hhmmss]{datetime}
\usetikzlibrary{calc}
\usetikzlibrary{decorations.pathreplacing}
\usetikzlibrary{patterns}
\usetikzlibrary{shadows}
\usetikzlibrary{arrows}
\usetikzlibrary{positioning}

\newlength\imageheight

\newcommand{\code}[1]{\texttt{#1}}
\newcommand*\tickmark{\textcolor{green!75!black}{\Checkmark}}

\usepackage[small,compact]{titlesec}

\newcommand\T{\rule{0pt}{2.6ex}}       
\newcommand\B{\rule[-1.2ex]{0pt}{0pt}} 

\title{Consus:  Taming the Paxi}
\author{\textnormal{Robert Escriva, Robbert van Renesse}\\
Computer Science Department, Cornell University
}

\begin{document}

\maketitle

\begin{abstract}
    Consus is a strictly serializable geo-replicated transactional key-value
    store.  The key contribution of Consus is a new commit protocol that reduces
    the cost of executing a transaction to three wide area message delays in the common case.
    Augmenting the commit protocol are multiple Paxos implementations optimized
    for different purposes.  Together the different implementations and
    optimizations comprise a cohesive system that provides low latency, high
    availability, and strong guarantees.  This paper describes the techniques
    implemented in the open source release of Consus, and lays the groundwork
    for evaluating Consus once the system implementation is sufficiently robust
    for a thorough evaluation.
\end{abstract}

\section{Introduction}
\label{sec:intro}

Geo-replication is a common feature among distributed storage systems.  A
geo-replicated system can withstand correlated failures up to and including
entire data centers, and may reduce latency for clients by directing them to
nearby data centers.  These systems differ from systems designed for a single
data center because they must account for latencies in the wide area that are
orders of magnitude larger than the latency for communication in a single data
center; a system designed for low latency settings will likely perform poorly if
geo-distributed.

The latency between geographically distinct locations forces systems to navigate
an inherent tradeoff between latency and fault tolerance.  Systems may make an
operation withstand a complete data center failure by incurring the latency cost
to propagate it to other data centers before reporting that the operation has
finished.  On the other side of the tradeoff, systems may avoid the latency cost
by reporting that an operation is complete before it propagates to other data
centers---at the risk that the operation is lost in a failure and never takes
effect.  An optimal point in this tradeoff would uphold a desired fault
tolerance guarantee while minimizing latency.

This paper introduces Consus\footnotemark{}, a geo-replicated key-value store
that supports strictly serializable cross-key transactions and executes
transactions with three wide area message delays in the common case.  It is easy
to see why avoiding latency is desirable: any latency incurred on the critical
path of a transaction directly impacts the performance of the application built
on top.  The decision to uphold strong guarantees is more a matter of taste; in
practice, organizations building on eventually consistent systems teach
developers anti-patterns to avoid or build special-purpose storage systems for
apps that are sensitive to consistency anomalies~\cite{facebook-consistency}.

\footnotetext{In Roman times, grain was essential to life and Consus served as
the protector of grain;  through its similarity to the word consilium, Consus
became associated with secret conferences.  In Modern times, Consus is very
similar to the word consensus; not coincidentally the former uses the latter to
serve as the protector of data.}

The core idea that enables Consus to reduce inter-data center latency is a
commit protocol based upon generalized consensus~\cite{gencon}.  Consus defers
all inter-data center communication to the commit protocol---leaving the commit
protocol to both globally replicate transactions and decide their outcomes.  The
commit protocol distributes the decision making process across all data centers
and typically completes in three inter-data center message delays.  Simply
sending a message to a remote data center and receiving acknowledgement of its
receipt---the bare minimum necessary to tolerate a failure---requires two such
delays.  Conventional commit protocols, such as 2-phase commit or Paxos, choose
a single ensemble member to aggregate and disseminate information and
communicating with this distinguished member necessarily incurs multiple
inter-data center delays.  Because Consus avoids a distinguished ensemble
member, Consus commits with 33\% less latency than other protocols, and incurs
one more delay than the minimum necessary to uphold any degree of fault
tolerance.

Consus' design makes a concerted effort to build on existing
protocols---primarily Paxos---to provide a principled argument for the system's
correctness.  Simply reusing an existing implementation of multi-Paxos would
suffice to uphold the safety guarantees of Consus and its commit protocol;
however, we will see that a generic Paxos implementation introduces latency and
failure sensitivities that negatively impact performance.  To overcome these
limitations, Consus uses multiple optimizations to Paxos that specialize Paxos
to the task at hand, rather than treating it as a black-box component.  Each
Paxos implementation relies upon a different set of architectural constraints
and protocol optimizations in order to decrease latency or improve availability
without sacrificing the safety guarantees made by Paxos.

Overall, this paper makes three contributions.  First and foremost, this paper
presents a new commit protocol which reduces the inter-data center communication
required to commit a transaction across multiple data centers.  Second, it
describes the optimized Paxos implementations within Consus outlining both the
rationale behind each optimization and the way in which it differs from generic
algorithms.  Finally, this paper describes the future direction of the project.

The rest of this paper is as follows:  Section~\ref{sec:design} lays out the
design of Consus, the inter-data center commit protocol, and the intra-data
center structure.  Section~\ref{sec:paxos} describes optimized or modified Paxos
protocols used within the design of Consus.  Section~\ref{sec:impl} describes
the state of the Consus implementation.  Section~\ref{sec:relwork} puts Consus
into context with related work.  Section~\ref{sec:discuss} discusses future work
on Consus and its publication.  The paper concludes with
Section~\ref{sec:concl}.

\section{Design}
\label{sec:design}

\begin{figure}[t]
    \centering
    \begin{tikzpicture}
\pgfdeclareimage[width=.95\linewidth]{usa}{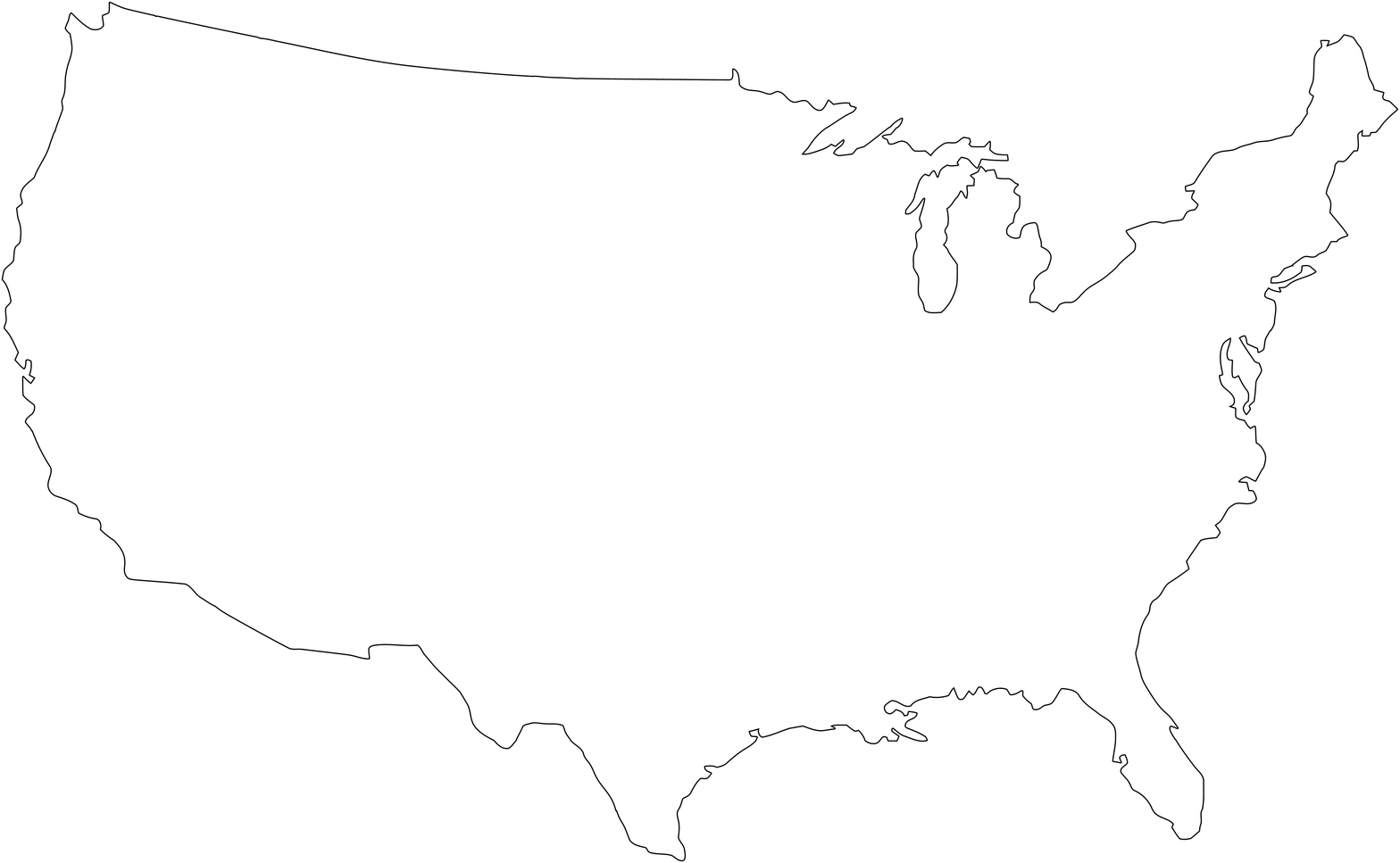};
\settoheight{\imageheight}{\pgfuseimage{usa}};
\pgfdeclareimage[height=1cm]{serv}{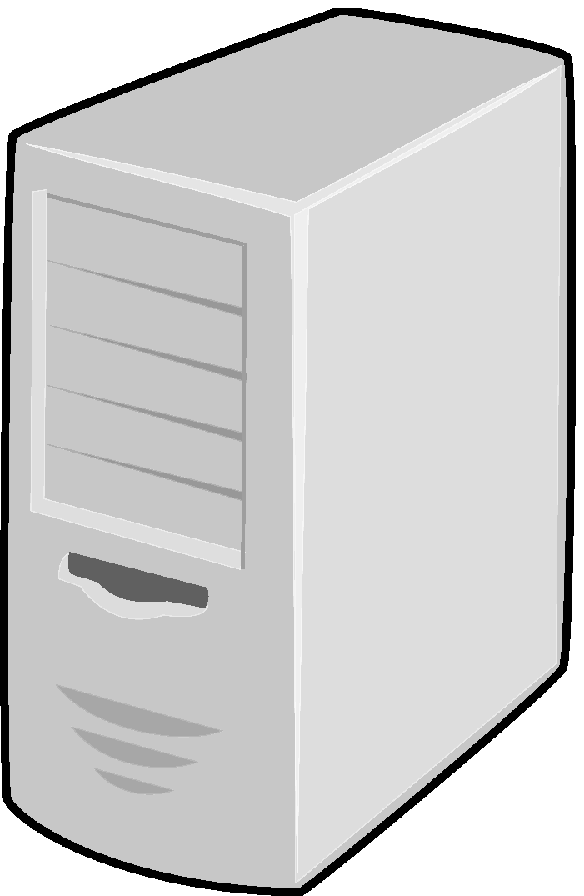};
\settoheight{\imageheight}{\pgfuseimage{serv}};
\pgfdeclareimage[height=1cm]{dbserv}{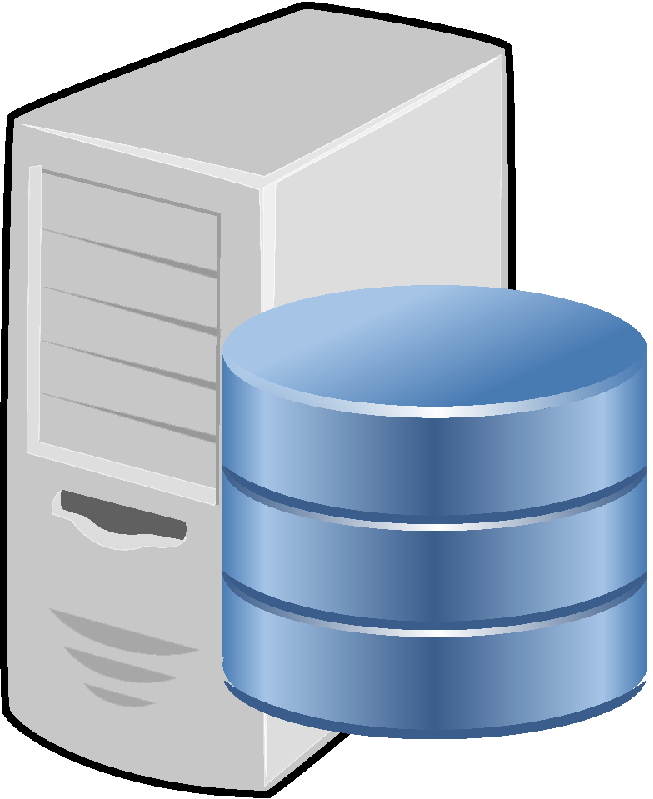};
\settoheight{\imageheight}{\pgfuseimage{dbserv}};
\pgfdeclareimage[height=1cm]{wwwserv}{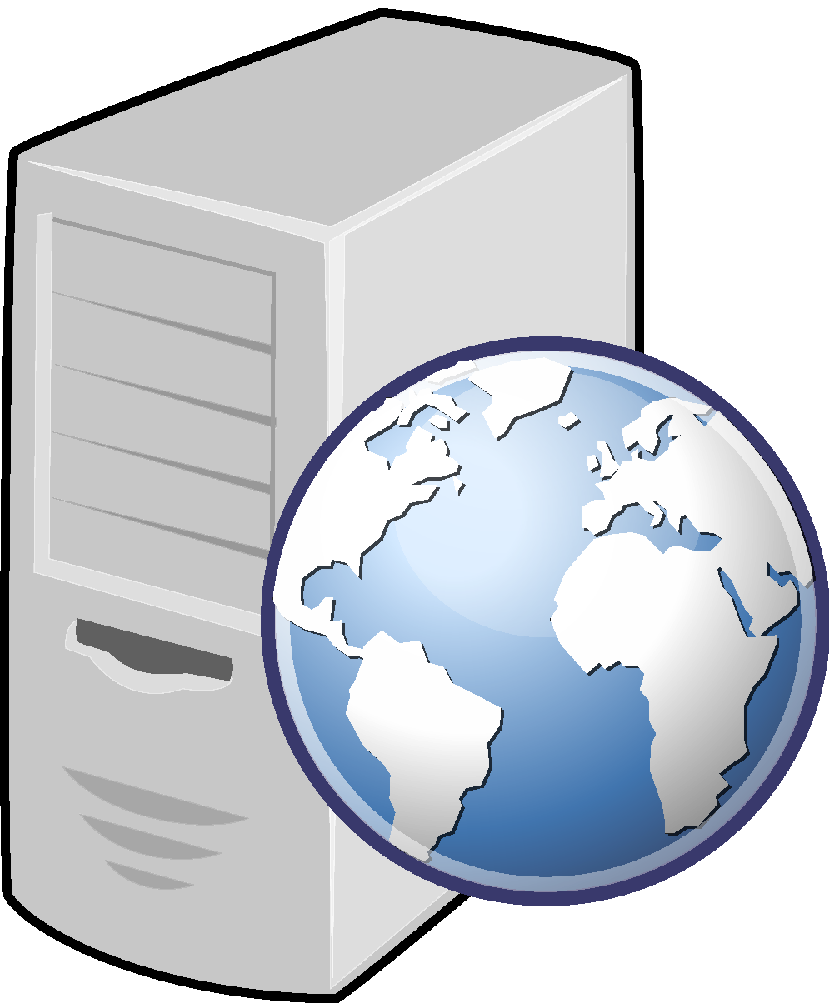};
\settoheight{\imageheight}{\pgfuseimage{wwwserv}};
\pgfdeclareimage[height=1cm]{client}{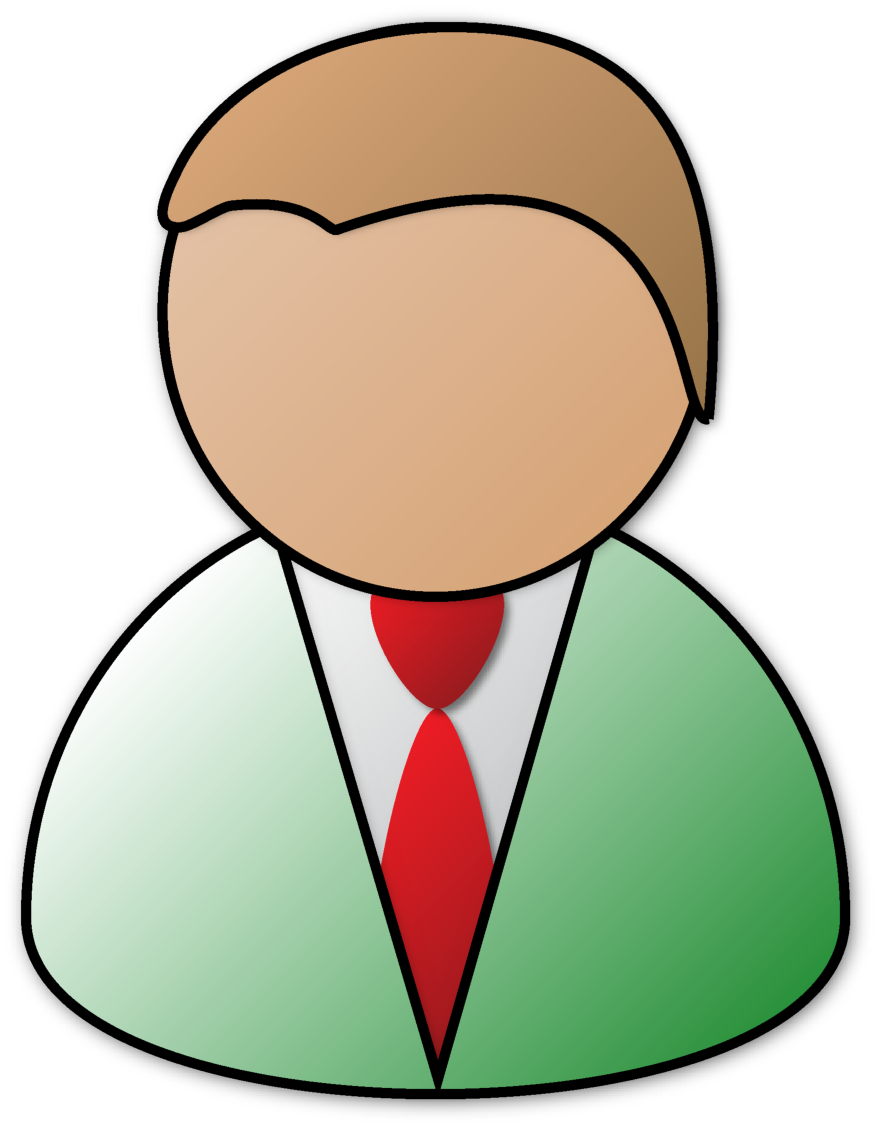};
\settoheight{\imageheight}{\pgfuseimage{client}};

\node[align=center] (usa) {\pgfuseimage{usa}};

\node (virginia) at (8em, -.75em) {\pgfuseimage{dbserv}};
\node (california) at (-10em, 0) {\pgfuseimage{dbserv}};
\node (oregon) at (-8em, 4.75em) {\pgfuseimage{dbserv}};

\draw[<->,line width=1pt] (virginia) -- (california) node [midway,sloped,color=black,fill=white] {Commit};
\draw[<->,line width=1pt] (oregon) -- (virginia) node [midway,sloped,color=black,fill=white] {Commit};
\draw[<->,line width=1pt] (california) -- (oregon);

\node[rectangle,draw=black,dotted,line width=1pt,minimum width=1cm,minimum height=1.25cm] (micro) at (virginia) {};
\node[below of=usa,node distance=15em,rectangle,draw=black,dotted,line width=1pt,minimum width=.98\linewidth,minimum height=13em] (macro) {};

\draw[dashed] (micro.south west) -- (macro.north west);
\draw[dashed] (micro.south east) -- (macro.north east);

\node[below of=usa,node distance=19em] (kvscenter) {$\cdots$};
\node[left of=kvscenter,node distance=3em] {\pgfuseimage{dbserv}};
\node[left of=kvscenter,node distance=6em] {\pgfuseimage{dbserv}};
\node[left of=kvscenter,node distance=9em] {\pgfuseimage{dbserv}};
\node[right of=kvscenter,node distance=3em] {\pgfuseimage{dbserv}};
\node[right of=kvscenter,node distance=6em] {\pgfuseimage{dbserv}};
\node[right of=kvscenter,node distance=9em] {\pgfuseimage{dbserv}};
\draw[dotted,line width=0.5pt]
    ($(kvscenter) + (10.5em, 1.75em)$) --
    ($(kvscenter) + (10.5em, -1.75em)$) --
    ($(kvscenter) + (-10.5em, -1.75em)$) --
    ($(kvscenter) + (-10.5em, 1.75em)$) --
    ($(kvscenter) + (10.5em, 1.75em)$);
\draw[dotted,line width=2pt]
    ($(kvscenter) + (-10.5em, 2.25em)$) --
    ($(kvscenter) + (10.5em, 2.25em)$)
    node [midway,color=black,fill=white] {Key Value Storage};

\node (txcenter) at ($(kvscenter) + (-3em, 7em)$) {$\cdots$};
\node[left of=txcenter,node distance=3em] {\pgfuseimage{serv}};
\node[left of=txcenter,node distance=6em] {\pgfuseimage{serv}};
\node[right of=txcenter,node distance=3em] {\pgfuseimage{serv}};
\draw[dotted,line width=0.5pt]
    ($(txcenter) + (4.5em, 1.75em)$) --
    ($(txcenter) + (4.5em, -1.75em)$) --
    ($(txcenter) + (-7.5em, -1.75em)$) --
    ($(txcenter) + (-7.5em, 1.75em)$) --
    ($(txcenter) + (4.5em, 1.75em)$);
\draw[dotted,line width=2pt]
    ($(txcenter) + (-7.5em, 2.25em)$) --
    ($(txcenter) + (4.5em, 2.25em)$)
    node [midway,color=black,fill=white] (txlabel) {Transaction Manager};

\node (commitcenter) at ($(kvscenter) + (8em, 7em)$) {\pgfuseimage{wwwserv}};
\draw[dotted,line width=0.5pt]
    ($(commitcenter) + (1.5em, 1.75em)$) --
    ($(commitcenter) + (1.5em, -1.75em)$) --
    ($(commitcenter) + (-1.5em, -1.75em)$) --
    ($(commitcenter) + (-1.5em, 1.75em)$) --
    ($(commitcenter) + (1.5em, 1.75em)$);
\draw[dotted,line width=2pt]
    ($(commitcenter) + (-2.25em, -1.85em)$) --
    ($(commitcenter) + (-2.25em, 2.25em)$) --
    node [midway,color=black,fill=white] (commitlabel) {Commit}
    ($(commitcenter) + (2.5em, 2.25em)$);

\draw[-implies,double distance=3pt]
    ($(txcenter) + (4.5em, 1em)$) -- node[midway,color=black,fill=white] {\footnotesize$T_x$ log}
    ($(commitcenter) + (-2.25em, 1em)$);

\draw[-implies,double distance=3pt]
    ($(commitcenter) + (-2.25em, -1em)$) -- node[midway,color=black,fill=white] {\footnotesize\textcolor{green!75!black}{\Checkmark}\ \footnotesize\textcolor{red!75!black}{\XSolidBrush}}
    ($(txcenter) + (4.5em, -1em)$);

\draw[implies-,double distance=3pt]
    ($(kvscenter) + (-8.5em, 2.25em)$) -- node[midway,sloped,color=black,fill=white] {\footnotesize{R}}
    ($(txcenter) + (-5.5em, -1.75em)$);

\draw[implies-,double distance=3pt]
    ($(kvscenter) + (-6.5em, 2.25em)$) -- node[midway,sloped,color=black,fill=white] {\footnotesize{W}}
    ($(txcenter) + (-3.5em, -1.75em)$);

\node[rectangle,above of=commitlabel,node distance=4em,draw=black,color=black,fill=white,dotted,line width=1pt] (other) {Other DCs};
\draw[implies-implies,double distance=3pt] ($(commitlabel) + (0, 0.25em)$) -- (other);

\node[above of=txlabel,node distance=4em] (user) {\pgfuseimage{client}};
\draw[implies-implies,double distance=3pt] ($(txlabel) + (0, 0.25em)$) -- (user);
\end{tikzpicture}
    \caption{A typical Consus deployment spanning three data centers.  Each data
    center resides in a single geographic region and is comprised of the commit
    protocol, a transaction manager, and key-value storage.  The commit protocol
    serves as the sole method of communication between the data centers.}
    \label{fig:arch}
\end{figure}

Consus uses well defined abstractions and builds upon proven protocols in order
to constrain complexity.  At the global scale, Consus treats each data center as
a singular entity and runs a commit protocol across these entities.  Internally
each data center is not a singular entity, but a cluster providing its own
partitioning and fault tolerance guarantees.  Each cluster is further subdivided
into a component for managing transaction execution, a component for storing the
key-value pairs, and a component for executing the commit protocol.  The commit
protocol serves as the singular point for inter-data center communication and
communicates with the local storage through a narrowly defined interface;
consequently, it is agnostic to the internals of the transaction manager or
key-value store.  Figure~\ref{fig:arch} summarizes this architecture.

Consus' design has a realistic set of assumptions.  It assumes that networks are
asynchronous and that servers may experience crash or omission failures.
Servers are assumed to recover all state from durable storage that was
acknowledged as durable.  Additionally. Consus assumes that there is some means
by which servers that permanently fail will eventually be tagged as failed and
removed from the cluster in order to restore the fault tolerance guarantees of
Paxos.  This mechanism need not be timely or accurate; a slow detection leaves
the cluster available, but with slightly weaker fault tolerance, while an
erroneous detection will be treated like any other crash failure.

\subsection{Commit Protocol}
\label{sec:inter}

The Consus commit protocol handles the task of executing a transaction across
multiple data centers.  The protocol takes as input a transaction from one data
center, replays it in other data centers, and outputs a decision to commit or
abort the transaction.  It is intentionally constrained to focus solely on
committing or aborting a transaction leaving all transaction execution and
concurrency control considerations to other components.

Consus decides the outcome of a transaction in three logically distinct phases.
In the first phase, a single data center executes the transaction; if the
transaction executes to completion, it is sent to other data centers alongside
sufficient information to determine that these data centers' executions match
the execution in the original data center.  In the second phase, each data
center broadcasts the result of its own execution---whether it was able to
reproduce the original execution---to all other data centers.  In the final
phase, the data centers feed these results to an instance of Generalized Paxos
that allows all data centers to learn the transaction's outcome.

It is the information learned during the final phase, combined with the
flexibility of Generalized Paxos, that enables the commit protocol's efficiency
improvements.  In 2-phase commit~\cite{2pc} and similar
protocols~\cite{3pc,2pcpaxos} the decision to commit is placed in an entity
called a coordinator.  Regardless of the coordinator's construction its
fundamental purpose is to aggregate and disseminate information between
participants in the protocol.  This necessarily introduces at least one message
delay for the coordinator to learn any information, and at least two message
delays before the other participants learn any information.  As we shall see as
we explore the full Consus commit protocol, all participants are able to learn
the requisite information---and be assured that all other participants will
learn the same information---in just one message delay during phase three.

\subsubsection*{Phase One}

Phase one of the commit protocol executes a transaction in one data center and
then re-executes it in the remaining data centers.  A re-execution is deemed
successful if and only if committing the transaction would leave the underlying
data affected by the transaction in a state that is indistinguishable from the
state of the same data in the initial data center.  In this way, the
re-execution preserves the original execution, and ensures all data centers
present a consistent view of the data.

Re-execution may fail for a variety of environmental or workload reasons.  For
example, re-executions may diverge when concurrently executing transactions
operate on the same data.  The process executing a transaction informs the
commit protocol when a re-execution diverges.  This enables each data center
participating in the commit protocol to use information about its own execution
in subsequent stages.

At the end of phase one, a data center knows its own (re-)execution outcome,
and, if the outcome is successful, is willing to hold the transaction ready to
commit until the protocol finishes.  The commit protocol requires that the
transaction manager guarantee the transaction can commit if the protocol outputs
a commit decision.

\subsubsection*{Phase Two}

Phase two of the commit protocol consists of each data center independently
broadcasting the result of its phase one execution to all other data centers.
It is not necessary for other data centers to finish---or have even
started---executing a transaction before receiving the phase two message from
another data center as long as the message's receipt is durably recorded.

There is no special insight to phase two because its purpose is exactly what it
seems:  Phase 2 informs most data center about most other data center's
executions.  Because failure is inevitable, it is impossible for every data
center to know about every other data center, nor whether any broadcasts were
received.  For this reason, while phase two is presented as a logically discrete
action, even completely operational data centers will continually broadcast
their execution to other data centers until the protocol runs to completion.

\subsubsection*{Phase Three}

Phase three of the commit protocol enables data centers to combine the
information broadcast by phase two into a single result learned by all data
centers.  Whereas phase two disseminates the outcome of the executions, phase
three aggregates them in a durable manner and ensures that all data centers
agree upon the aggregated value.

The value learned by data centers in phase three is a set of \code{commit} or
\code{abort} results from the data centers involved in the transaction.  This
set of results is maintained by an instance of Generalized Paxos.  As individual
data centers learn the set of results from the Generalized Paxos protocol, the
data centers count the results and decide whether to commit or abort the
transaction.

The Generalized Paxos protocol learns a partially ordered set (poset) of values.
In phase three, these values are the phase two results, and there exists no
order across these results.  The primary contribution of Generalized Paxos is to
provide a fast path by which acceptors can extend the accepted value poset by
directly adding elements, so long as each acceptor has the same partial order
across elements.  Because phase two results are inherently unordered, acceptors
can always propose these results and remain on the fast path of Generalized
Paxos.

The structure of phase three allows each data center's acceptor to independently
accept a value with sufficient information to decide to commit or abort.  The
data centers can then broadcast this accepted value with a single Paxos Phase 2B
message to the other data centers.  Once each data center receives a quorum of
these Phase 2B messages, it can independently learn the poset of results without
consulting other data centers.

In summary, at the beginning of phase three, each data center has the result
from every other data center.  Each data center proposes to its own local
acceptor these values and then broadcasts its acceptor's state.  This third
broadcast is the third message delay in the commit protocol.  Upon receiving the
majority of these broadcasts, every data center can calculate the learned value
of these accepted values without any single data center coordinating the learned
value.

\subsubsection*{Avoiding Deadlock}

\begin{figure}[t]
    \centering
    \begin{tikzpicture}
\pgfdeclareimage[width=.95\linewidth]{usa}{USA.ps};
\settoheight{\imageheight}{\pgfuseimage{usa}};
\pgfdeclareimage[height=1cm]{serv}{server.ps};
\settoheight{\imageheight}{\pgfuseimage{serv}};
\pgfdeclareimage[height=1cm]{dbserv}{database_server.ps};
\settoheight{\imageheight}{\pgfuseimage{dbserv}};
\pgfdeclareimage[height=1cm]{wwwserv}{web_server.ps};
\settoheight{\imageheight}{\pgfuseimage{wwwserv}};

\node[align=center] (usa) {\pgfuseimage{usa}};

\node (virginia) at (8em, -.75em) {\pgfuseimage{dbserv}};
\node (california) at (-10em, 0) {\pgfuseimage{dbserv}};
\node (oregon) at (-8em, 4.75em) {\pgfuseimage{dbserv}};

\node[circle,color=white,fill=red,draw=white,inner sep=1pt,minimum width=0.1em] (v1) at (virginia) {\small 1};
\node[right of=v1,node distance=0.9em,circle,color=white,fill=gray,draw=white,inner sep=1pt,minimum width=0.1em] (v3) {\small 3};
\node[right of=v3,node distance=0.9em,circle,color=white,fill=gray,draw=white,inner sep=1pt,minimum width=0.1em] (v2) {\small 2};

\node[circle,color=white,fill=red,draw=white,inner sep=1pt,minimum width=0.1em] (c2) at (california) {\small 2};
\node[right of=c2,node distance=0.9em,circle,color=white,fill=gray,draw=white,inner sep=1pt,minimum width=0.1em] (c1) {\small 1};
\node[right of=c1,node distance=0.9em,circle,color=white,fill=gray,draw=white,inner sep=1pt,minimum width=0.1em] (c3) {\small 3};

\node[circle,color=white,fill=red,draw=white,inner sep=1pt,minimum width=0.1em] (o3) at (oregon) {\small 3};
\node[right of=o3,node distance=0.9em,circle,color=white,fill=gray,draw=white,inner sep=1pt,minimum width=0.1em] (o2) {\small 2};
\node[right of=o2,node distance=0.9em,circle,color=white,fill=gray,draw=white,inner sep=1pt,minimum width=0.1em] (o1) {\small 1};

\end{tikzpicture}
    \caption{An example deadlock where three independent transactions block each
    other from making progress.  The commit protocol can take upcalls from the
    transaction execution component to avoid deadlock.}
    \label{fig:deadlock}
\end{figure}
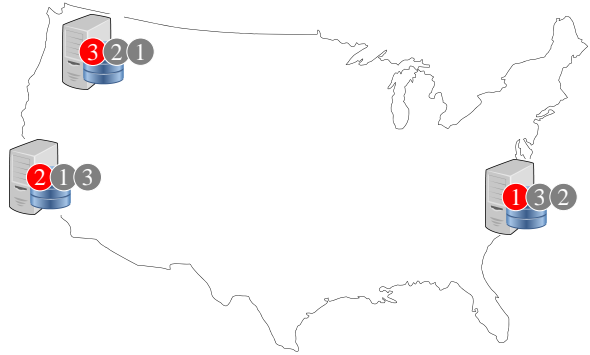

The invariants that the Consus commit protocol imposes on the rest of the system
make it possible for transactions to deadlock.  Specifically, the constraint
that a transaction remain committable until the commit protocol returns a
decision will introduce behavior analogous the classic problem of deadlock in a
lock-based database system.  Figure~\ref{fig:deadlock} shows an example deadlock
arising from three data centers each touching the same data.

To prevent deadlock, the commit protocol accepts upcalls from the transaction
execution component that indicate when a transaction may be potentially
deadlocked.  Such an upcall signals to the commit protocol that the transaction
might not ever commit and the protocol should act to avoid the deadlock.  For
some transactions, the transaction's outcome may already be decided by the time
the upcall reaches the commit protocol.  These transactions require no special
consideration.

The commit protocol will attempt to abort transactions that have no outcome at
the time of the upcall.  This process is complicated by the fact that an upcall
may be generated within any data center, but all data centers must uniformly
agree on a transaction's outcome.  Consequently, all data centers must also
agree to act to abort a transaction in response to a potential deadlock.

In response to deadlock, data centers may atomically signal their intent to
retract a previously-recorded commit result and replace it with an abort result.
To do this, the data center proposes the upcall to the Generalized Paxos
instance of phase three of the commit protocol.  Other data centers can then
negate the previous \code{commit} result and count it as an \code{abort} result
instead.

In order to ensure that a retraction is treated the same by all data centers,
retractions are totally ordered with respect to all other elements in the
learned poset.  This partial ordering guarantees that data centers will not
diverge when counting results.  The data center tallies results in accordance
with the partial order present in the set.  If the commit total exceeds a quorum
before a retraction is processed, the retraction is ignored; otherwise, the
total shifts toward aborting the transaction.  The partial order induced on the
set ensures that results may be seen in any order, but the total \code{commit}
and \code{abort} results encountered before a retraction are the same in all
learned values.  Similarly, because retractions are totally ordered all data
centers will see the retractions in the same order.

The deadlock avoidance algorithm may force Generalized Paxos protocol to fall
back to a classic Paxos round to resolve conflicts.  These conflicts occur when
the partial order at one acceptor differs from the partial order at another
acceptor.  Generalized Paxos will fall back to one or more rounds of the
traditional Paxos protocol to resolve these conflicts.  Because these additional
rounds of Paxos take place in the wide area, they increase the number of message
delays a transaction will encounter---however this cost is incurred only when
data centers are attempting to abort a transaction due to deadlock.

Heuristics can be used to propose retractions to Generalized Paxos in a way that
is unlikely to generate conflicts.  For example, each data center may delay
proposing any retractions to its local acceptor until the set of results it has
previously proposed combined with the unproposed retractions would yield an
abort outcome.  The data center could then propose the retractions in a
pre-determined order that ensures that all data centers propose retractions in
the same order.  This is the simplest heuristic one could employ to keep
Generalized Paxos from reverting to Classic Paxos; it is certainly worth
investigating other heuristics alongside an investigation of ways to reduce the
likelihood of abort upcalls in the underlying transaction execution engine.

\subsubsection*{Data Center Failure}

Thus far, we have explored the commit protocol without regard to data center
failure.  While the protocol is resilient to a minority of data center failures,
additional considerations are necessary to make the protocol return to a steady
state after a data center recovers from failure.  Specifically, any transactions
executed while a data center is unavailable will not be propagated to the data
center by the commit protocol; an additional mechanism is necessary to propagate
any missed transactions.

A simple, but inefficient, approach is to have every data center continually
synchronize state with other data centers.  This ensures that any data committed
in a majority of data centers will eventually propagate to all other data
centers.  The downside to this technique is that it requires background
synchronization and any transaction that relies upon data that has yet to
propagate will abort at all out-of-sync data centers.

In order to quickly bring data centers up to date and avoid background
communication, Consus also uses the information embedded in the phase one
execution log to bring a data center up to date.  The execution log includes
information about every data item read during the transaction.  Because
transactions will only read committed data, it stands to reason that any read
performed within a transaction is reading data that must exist in the key-value
store.  Consequently, the transaction execution engine can turn a read for
missing data into an implicit write that restores missing data.  While this is
less efficient than reading the data, it incurs no latency waiting for the data
center to become up-to-date---the data center can immediately commit the
transaction.

Consus employs both of these techniques to recover from data center failure.
The former technique ensures that all data becomes replicated to all requisite
data centers, while the latter technique prevents a data center from appearing
unavailable between cross-data center synchronization events.

\subsection{Intra-Data Center Design}
\label{sec:intra}

Within each data center, Consus is divided into two distinct execution
components:  a transaction manager and key-value storage.  The transaction
manager presents the sole interface to the client and uses two-phase locking as
its concurrency control mechanism.  The key-value storage serves as the
component of record for each object stored within Consus and also stores the
locks used by the transaction manager.

\subsubsection*{Transaction Manager}

The transaction manager component durably records transactions during their
execution.  This ensures that all information about a transaction is recorded in
one location and not scattered about the cluster.  Upon failure of an entire
data center, this logically centralized location provides a direct means to
resume a transactions' execution without having to gather the information from
many disparate points around the cluster.

\begin{figure}[t]
    \centering
    \begin{tikzpicture}
\pgfdeclareimage[height=.5cm]{serv}{server.ps};
\settoheight{\imageheight}{\pgfuseimage{serv}};
\pgfdeclareimage[height=.5cm]{client}{business_person.ps};
\settoheight{\imageheight}{\pgfuseimage{client}};

\node (origin) {};

\node[right of=origin,node distance=0em] (cluster1) {};
\node[right of=origin,node distance=5em] (cluster2) {};
\node[right of=origin,node distance=10em] (cluster3) {};
\node[right of=origin,node distance=16em] (cluster4) {};

\draw[thick,dotted] (cluster1) circle (1.75em);
\foreach \name/\angle in {c1s1/0, c1s2/72, c1s3/144, c1s4/216, c1s5/288}
    \node (\name) at ($(cluster1) + (\angle:1.75em)$) {\pgfuseimage{serv}};

\draw[thick,dotted] (cluster2) circle (1.75em);
\foreach \name/\angle in {c2s1/0, c2s2/72, c2s3/144, c2s4/216, c2s5/288}
    \node (\name) at ($(cluster2) + (\angle:1.75em)$) {\pgfuseimage{serv}};

\draw[thick,dotted] (cluster3) circle (1.75em);
\foreach \name/\angle in {c3s1/0, c3s2/72, c3s3/144, c3s4/216, c3s5/288}
    \node (\name) at ($(cluster3) + (\angle:1.75em)$) {\pgfuseimage{serv}};

\draw[thick,dotted] (cluster4) circle (1.75em);
\foreach \name/\angle in {c4s1/0, c4s2/72, c4s3/144, c4s4/216, c4s5/288}
    \node (\name) at ($(cluster4) + (\angle:1.75em)$) {\pgfuseimage{serv}};

\draw[rectangle,draw=black,line width=1pt,dashed]
    ($(cluster1) + (-2.5em,-2.7em)$)
    -- ($(cluster3) + (2.7em,-2.7em)$)
    -- ($(cluster3) + (2.7em,2.7em)$)
    -- ($(cluster1) + (-2.5em,2.7em)$)
    -- ($(cluster1) + (-2.5em,-2.7em)$);

\draw[rectangle,draw=black,line width=1pt,dashed]
    ($(cluster4) + (-2.5em,-2.7em)$)
    -- ($(cluster4) + (2.7em,-2.7em)$)
    -- ($(cluster4) + (2.7em,2.7em)$)
    -- ($(cluster4) + (-2.5em,2.7em)$)
    -- ($(cluster4) + (-2.5em,-2.7em)$);

\node[above of=cluster4,node distance=6em] (overload) {};
\node[left of=overload,node distance=4em] (overclient1) {\pgfuseimage{client}};
\node[left of=overload,node distance=3em] (overclient2) {\pgfuseimage{client}};
\node[left of=overload,node distance=2em] (overclient3) {\pgfuseimage{client}};
\node[left of=overload,node distance=1em] (overclient4) {\pgfuseimage{client}};
\node[right of=overload,node distance=0em] (overclient5) {\pgfuseimage{client}};
\node[right of=overload,node distance=1em] (overclient6) {\pgfuseimage{client}};
\node[right of=overload,node distance=2em] (overclient7) {\pgfuseimage{client}};

\draw[->,thick] (overclient1) -- (c4s4);
\draw[->,thick] (overclient2) -- (c4s4);
\draw[->,thick] (overclient3) -- (c4s3);
\draw[->,thick] (overclient4) -- (c4s2);
\draw[->,thick] (overclient5) -- (c4s2);
\draw[->,thick] (overclient6) -- (c4s2);
\draw[->,thick] (overclient7) -- (c4s5);

\node[above of=cluster2,node distance=6em] (lightload) {};
\node[left of=lightload,node distance=5em] (lightclient1) {\pgfuseimage{client}};
\node[left of=lightload,node distance=1.67em] (lightclient2) {\pgfuseimage{client}};
\node[right of=lightload,node distance=1.67em] (lightclient3) {\pgfuseimage{client}};
\node[right of=lightload,node distance=5em] (lightclient4) {\pgfuseimage{client}};

\draw[->,thick] (lightclient1) -- (c1s3);
\draw[->,thick] (lightclient2) -- (c1s1);
\draw[->,thick] (lightclient2) -- (c2s5);
\draw[->,thick] (lightclient3) -- (c2s2);
\draw[->,thick] (lightclient4) -- (c2s1);

\draw[rectangle,draw=black,dotted,line width=1pt]
       ($(lightclient1) + (-.85em,-1em)$)
    -- ($(lightclient4) +  (.85em,-1em)$)
    -- ($(lightclient4) +  (.85em, 1.25em)$)
    -- node[midway,fill=white,inner sep=0pt] {\footnotesize{Application 1}}
       ($(lightclient1) + (-.85em, 1.25em)$)
    -- ($(lightclient1) + (-.85em,-1em)$);

\draw[rectangle,draw=black,dotted,line width=1pt]
       ($(overclient1) + (-.85em,-1em)$)
    -- ($(overclient7) +  (.85em,-1em)$)
    -- ($(overclient7) +  (.85em, 1.25em)$)
    -- node[midway,fill=white,inner sep=0pt] {\footnotesize{Application 2}}
       ($(overclient1) + (-.85em, 1.25em)$)
    -- ($(overclient1) + (-.85em,-1em)$);

\end{tikzpicture}
    \caption{The transaction manager component.  Transactions are sharded across
    Paxos groups that replicate each transaction as a state machine.  Different
    applications may be directed to different state machines in order to provide
    performance isolation at the transaction level.}
    \label{fig:txman}
\end{figure}
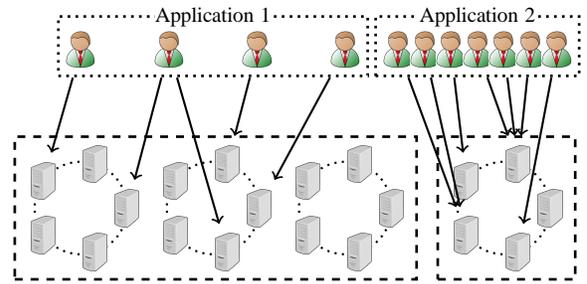

For scalability and fault tolerance, the transaction manager is partitioned
across multiple Paxos groups.  Each transaction executes as a replicated
state machine at exactly one of these Paxos groups.  Consequently, the tier can
be scaled to accommodate more transactions by adding additional servers and
Paxos groups.  With each additional Paxos group comes the ability to log more
operations to disk.  For workloads with low contention, this leads to a direct
increase in performance.  Figure~\ref{fig:txman} summarizes this architecture.

Transactions execute as a replicated state machine at a single Paxos group.
Clients submit begin, read, write, and commit operations to the group, which
then durably records and agrees upon the sequence of events issued by the
client.  The group interacts with the rest of the cluster on behalf of the
client, acquiring locks, proxying reads, buffering writes, and releasing locks.

When the client commits the transaction, the Paxos group will hand the entire
record of the transaction to the commit protocol.  If the transaction commits,
the transaction manager will flush the buffered writes to the key-value store
before releasing locks, while immediately acknowledging the commit to the
client.  Thus, even when the client fails, the transaction manager group may
push a transaction to completion, or abort the transaction and clean up any
state affected by the transaction.

One benefit of this sharded and replicated structure that is not immediately
obvious is the ability to isolate multiple tenants' transaction managers by
directing different tenants or applications to different Paxos groups.  This
isolation at the hardware level makes it impossible for one overly aggressive
application to affect the performance of other, possibly higher priority,
applications' transaction execution.  This isolation only goes as far as
isolation at the transaction manager level.  Transactions that touch the same
data are not isolated because of contention at the key-value store.

\subsubsection*{Key-Value Storage}

The key-value store maintains a partitioned sorted map from bytestring keys to
stored objects.  It internally handles replication and partitioning of the data
across multiple storage servers to enable the cluster to scale to many petabytes
in size.  By keeping the details of the key-value store's replication and
partitioning mechanisms internal to the key-value component, the component's
interface may be simplified to a small number of well-defined RPCs that may be
issued to any server in the key-value store.  This decision is in contrast to
systems like HyperDex~\cite{hyperdex} that maintain logic in the client library
for routing requests to servers, and re-routing or failing requests if the
server configuration changes.  Although the design in Consus potentially incurs
an extra messaging hop within the data center, it avoids having to distribute
the key-value mapping to clients of the key-value store; this distribution can
become increasingly expensive as the cluster grows in size, and the extra round
trip avoids ever incurring this cost.

Consus divides the key-value space into a constant number of partitions in order
to simplify the implementation and allow for global decisions during
repartitioning events.  By maintaining a constant number of partitions, Consus
can compactly represent a mapping from each server to a range of partitions
the server stores.  Replicas of the data are stored by the servers adjacent to
the data in the key space; this enables the servers to takeover serving a
partition with minimal movement of data across servers.  This design is in
contrast to systems which store fixed-size ranges of the key-space and maintain
a lookup table between keys and the range they map to.  Such a scheme can grow
to an unlimited number of constant-sized ranges and require that extra state be
maintained to track the ranges themselves.

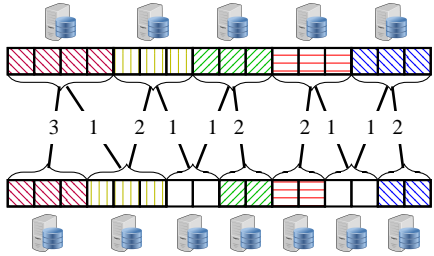
\begin{figure}[t]
    \centering
    \begin{tikzpicture}
\pgfdeclareimage[height=.5cm]{dbserv}{database_server.ps};
\settoheight{\imageheight}{\pgfuseimage{dbserv}};

\node (before) {};

\draw[thick,pattern=north west lines,pattern color=purple]
    ($(before) + (-8em,0em)$) rectangle ($(before) + (-4em,1em)$);
\draw[thick,pattern=vertical lines,pattern color=yellow!75!black]
    ($(before) + (-4em,0em)$) rectangle ($(before) + (-1em,1em)$);
\draw[thick,pattern=north east lines,pattern color=green!75!black]
    ($(before) + (-1em,0em)$) rectangle ($(before) + (2em,1em)$);
\draw[thick,pattern=horizontal lines,pattern color=red]
    ($(before) + (2em,0em)$)  rectangle ($(before) + (5em,1em)$);
\draw[thick,pattern=north west lines,pattern color=blue]
    ($(before) + (5em,0em)$)  rectangle ($(before) + (8em,1em)$);

\draw[thick] ($(before) + (-7em,0em)$) rectangle ($(before) + (-8em,1em)$);
\draw[thick] ($(before) + (-6em,0em)$) rectangle ($(before) + (-7em,1em)$);
\draw[thick] ($(before) + (-5em,0em)$) rectangle ($(before) + (-6em,1em)$);
\draw[thick] ($(before) + (-4em,0em)$) rectangle ($(before) + (-5em,1em)$);
\draw[thick] ($(before) + (-3em,0em)$) rectangle ($(before) + (-4em,1em)$);
\draw[thick] ($(before) + (-2em,0em)$) rectangle ($(before) + (-3em,1em)$);
\draw[thick] ($(before) + (-1em,0em)$) rectangle ($(before) + (-2em,1em)$);
\draw[thick] ($(before) + (0em,0em)$)  rectangle ($(before) + (-1em,1em)$);
\draw[thick] ($(before) + (0em,0em)$)  rectangle ($(before) + (1em,1em)$);
\draw[thick] ($(before) + (1em,0em)$)  rectangle ($(before) + (2em,1em)$);
\draw[thick] ($(before) + (2em,0em)$)  rectangle ($(before) + (3em,1em)$);
\draw[thick] ($(before) + (3em,0em)$)  rectangle ($(before) + (4em,1em)$);
\draw[thick] ($(before) + (4em,0em)$)  rectangle ($(before) + (5em,1em)$);
\draw[thick] ($(before) + (5em,0em)$)  rectangle ($(before) + (6em,1em)$);
\draw[thick] ($(before) + (6em,0em)$)  rectangle ($(before) + (7em,1em)$);
\draw[thick] ($(before) + (7em,0em)$)  rectangle ($(before) + (8em,1em)$);

\node[above of=before,node distance=2em] (beforeservs) {};
\node[left of=beforeservs,node distance=6em] {\pgfuseimage{dbserv}};
\node[left of=beforeservs,node distance=2.5em] {\pgfuseimage{dbserv}};
\node[right of=beforeservs,node distance=0.5em] {\pgfuseimage{dbserv}};
\node[right of=beforeservs,node distance=3.5em] {\pgfuseimage{dbserv}};
\node[right of=beforeservs,node distance=6.5em] {\pgfuseimage{dbserv}};

\draw[decorate,decoration={brace,amplitude=6pt},xshift=-4pt,yshift=0pt]
    ($(before) + (-4em,0em)$) -- node[black,midway,yshift=-1pt] (partb1) {}
    ($(before) + (-8em,0em)$);

\draw[decorate,decoration={brace,amplitude=6pt},xshift=-4pt,yshift=0pt]
    ($(before) + (-1em,0em)$) -- node[black,midway,yshift=-1pt] (partb2) {}
    ($(before) + (-4em,0em)$);

\draw[decorate,decoration={brace,amplitude=6pt},xshift=-4pt,yshift=0pt]
    ($(before) + (2em,0em)$) -- node[black,midway,yshift=-1pt] (partb3) {}
    ($(before) + (-1em,0em)$);

\draw[decorate,decoration={brace,amplitude=6pt},xshift=-4pt,yshift=0pt]
    ($(before) + (5em,0em)$) -- node[black,midway,yshift=-1pt] (partb4) {}
    ($(before) + (2em,0em)$);

\draw[decorate,decoration={brace,amplitude=6pt},xshift=-4pt,yshift=0pt]
    ($(before) + (8em,0em)$) -- node[black,midway,yshift=-1pt] (partb5) {}
    ($(before) + (5em,0em)$);


\node[below of=before,node distance=5em] (after) {};

\draw[thick,pattern=north west lines,pattern color=purple]
    ($(after) + (-8em,0em)$) rectangle ($(after) + (-5em,1em)$);
\draw[thick,pattern=vertical lines,pattern color=yellow!75!black]
    ($(after) + (-5em,0em)$) rectangle ($(after) + (-2em,1em)$);
\draw[thick,pattern=north east lines,pattern color=green!75!black]
    ($(after) + (0em,0em)$) rectangle ($(after) + (2em,1em)$);
\draw[thick,pattern=horizontal lines,pattern color=red]
    ($(after) + (2em,0em)$)  rectangle ($(after) + (4em,1em)$);
\draw[thick,pattern=north west lines,pattern color=blue]
    ($(after) + (6em,0em)$)  rectangle ($(after) + (8em,1em)$);

\draw[thick] ($(after) + (-7em,0em)$) rectangle ($(after) + (-8em,1em)$);
\draw[thick] ($(after) + (-6em,0em)$) rectangle ($(after) + (-7em,1em)$);
\draw[thick] ($(after) + (-5em,0em)$) rectangle ($(after) + (-6em,1em)$);
\draw[thick] ($(after) + (-4em,0em)$) rectangle ($(after) + (-5em,1em)$);
\draw[thick] ($(after) + (-3em,0em)$) rectangle ($(after) + (-4em,1em)$);
\draw[thick] ($(after) + (-2em,0em)$) rectangle ($(after) + (-3em,1em)$);
\draw[thick] ($(after) + (-1em,0em)$) rectangle ($(after) + (-2em,1em)$);
\draw[thick] ($(after) + (0em,0em)$)  rectangle ($(after) + (-1em,1em)$);
\draw[thick] ($(after) + (0em,0em)$)  rectangle ($(after) + (1em,1em)$);
\draw[thick] ($(after) + (1em,0em)$)  rectangle ($(after) + (2em,1em)$);
\draw[thick] ($(after) + (2em,0em)$)  rectangle ($(after) + (3em,1em)$);
\draw[thick] ($(after) + (3em,0em)$)  rectangle ($(after) + (4em,1em)$);
\draw[thick] ($(after) + (4em,0em)$)  rectangle ($(after) + (5em,1em)$);
\draw[thick] ($(after) + (5em,0em)$)  rectangle ($(after) + (6em,1em)$);
\draw[thick] ($(after) + (6em,0em)$)  rectangle ($(after) + (7em,1em)$);
\draw[thick] ($(after) + (7em,0em)$)  rectangle ($(after) + (8em,1em)$);

\node[below of=after,node distance=1em] (afterservs) {};
\node[left of=afterservs,node distance=6.5em] {\pgfuseimage{dbserv}};
\node[left of=afterservs,node distance=3.5em] {\pgfuseimage{dbserv}};
\node[left of=afterservs,node distance=1em] {\pgfuseimage{dbserv}};
\node[right of=afterservs,node distance=1em] {\pgfuseimage{dbserv}};
\node[right of=afterservs,node distance=3em] {\pgfuseimage{dbserv}};
\node[right of=afterservs,node distance=5em] {\pgfuseimage{dbserv}};
\node[right of=afterservs,node distance=7em] {\pgfuseimage{dbserv}};

\draw[decorate,decoration={brace,amplitude=6pt},xshift=-4pt,yshift=0pt]
    ($(after) + (-8em,1em)$) -- node[black,midway,yshift=1pt] (parta1) {}
    ($(after) + (-5em,1em)$);

\draw[decorate,decoration={brace,amplitude=6pt},xshift=-4pt,yshift=0pt]
    ($(after) + (-5em,1em)$) -- node[black,midway,yshift=1pt] (parta2) {}
    ($(after) + (-2em,1em)$);

\draw[decorate,decoration={brace,amplitude=6pt},xshift=-4pt,yshift=0pt]
    ($(after) + (-2em,1em)$) -- node[black,midway,yshift=1pt] (parta3) {}
    ($(after) + (0em,1em)$);

\draw[decorate,decoration={brace,amplitude=6pt},xshift=-4pt,yshift=0pt]
    ($(after) + (0em,1em)$) -- node[black,midway,yshift=1pt] (parta4) {}
    ($(after) + (2em,1em)$);

\draw[decorate,decoration={brace,amplitude=6pt},xshift=-4pt,yshift=0pt]
    ($(after) + (2em,1em)$) -- node[black,midway,yshift=1pt] (parta5) {}
    ($(after) + (4em,1em)$);

\draw[decorate,decoration={brace,amplitude=6pt},xshift=-4pt,yshift=0pt]
    ($(after) + (4em,1em)$) -- node[black,midway,yshift=1pt] (parta6) {}
    ($(after) + (6em,1em)$);

\draw[decorate,decoration={brace,amplitude=6pt},xshift=-4pt,yshift=0pt]
    ($(after) + (6em,1em)$) -- node[black,midway,yshift=1pt] (parta7) {}
    ($(after) + (8em,1em)$);


\draw[-,line width=1pt] (partb1) -- node[midway,black,fill=white] {\footnotesize{3}} (parta1);
\draw[-,line width=1pt] (partb1) -- node[midway,black,fill=white] {\footnotesize{1}} (parta2);

\draw[-,line width=1pt] (partb2) -- node[midway,black,fill=white] {\footnotesize{2}} (parta2);
\draw[-,line width=1pt] (partb2) -- node[midway,black,fill=white] {\footnotesize{1}} (parta3);

\draw[-,line width=1pt] (partb3) -- node[midway,black,fill=white] {\footnotesize{1}} (parta3);
\draw[-,line width=1pt] (partb3) -- node[midway,black,fill=white] {\footnotesize{2}} (parta4);

\draw[-,line width=1pt] (partb4) -- node[midway,black,fill=white] {\footnotesize{2}} (parta5);
\draw[-,line width=1pt] (partb4) -- node[midway,black,fill=white] {\footnotesize{1}} (parta6);

\draw[-,line width=1pt] (partb5) -- node[midway,black,fill=white] {\footnotesize{1}} (parta6);
\draw[-,line width=1pt] (partb5) -- node[midway,black,fill=white] {\footnotesize{2}} (parta7);

\end{tikzpicture}
    \caption{A key-value store mapping scaling from 5 nodes to 7.  The overlap
    between the old and new mapping is used to calculate a preference matching
    servers to positions in the new mapping.  The stable marriage algorithm then
    determines which ranges of partitions servers will adopt.  Ranges left empty
    by the stable marriage algorithm are assigned to new servers.}
    \label{fig:marriage}
\end{figure}

When servers join or depart the cluster, a single fault tolerant process within
the cluster determines a globally optimal rebalanced form of the cluster and
issues this new mapping to servers within the cluster.  The process uses a
variant of the stable marriage algorithm~\cite{stablemarriage} to rebalance the
cluster.  The algorithm creates a new ideal mapping to the key-value storage
nodes that divides the constant number of partitions in the cluster into
contiguous ranges of partitions, each of which will be assigned to a single
node.  It then assigns a preference each server has for each of these ranges by
counting the number of partitions the server has that fall within the range.
This preference between servers and partitions serves as input to the stable
marriage algorithm, which will align each server with a range of partitions with
a guarantee analogous to a stable marriage in the original algorithm.
Figure~\ref{fig:marriage} shows an 8-partition cluster growing from five nodes
to seven, the preferences between the old nodes and new partitions, and how new
nodes are distributed to the unassigned partitions.

To maximize the effectiveness of this assignment, the key-value store
incrementally changes from one assignment to the next.  Servers will
incrementally adopt their assigned ranges, one partition at a time to ensure
that at all times they are assigned to a contiguous range of partitions.  This
guarantees that any incremental assignment between two mappings can also serve
as an input to the global optimization algorithm.  Thus, work performed
migrating from one configuration to the next will can be reused---even when the
cluster dramatically changes in structure.

One open question regarding the stable marriage algorithm is the extent to which
replication should be considered.  The current design considers only a singular
server for each partition, but could be adapted to consider replication when
computing the weighting.  In practice, this would likely produce a very similar
result, but would be complex when disparate key spaces hosted on the key-value
store specify different replication factors.

\section{Paxos Optimizations}
\label{sec:paxos}

Consus uses multiple optimizations to Paxos to provide better performance than
off-the-shelf Paxos implementations.  In this section we will explore these
optimizations and examine how they can improve on an off-the-shelf Paxos
protocol.  For each technique described, we will look at why the technique
retains the safety that Consus requires of Paxos and, where relevant, why the
technique is not generally applicable to all Paxos implementations.

\subsection{Avoiding Paxos}

The most elementary optimization is to avoid using Paxos entirely.  The
transaction state machine in the transaction manager was originally implemented
using a standard Multi-Paxos protocol.  Because of the unique structure of the
transaction state machine, namely that there is a single source of all
proposals, it was easy to completely avoid using Paxos with no loss of safety.

Because there is a single proposer, sequencing the operations does not require
consensus---the single proposer has already sequenced the operations.  The
members of the ensemble need only durably log this sequence.  In an
off-the-shelf Multi-Paxos library, the single client would send all operations
through a single member that would propose the operations to the cluster.

The reason that Consus is able to avoid Paxos entirely in this particular
instance is because the client and the transaction have a shared fate.  If the
client dies, there cannot be any more proposals issued to the cluster.
Therefore when the client dies, the transaction log will not grow further.  If a
client dies, or takes too long to issue additional operations to a transaction,
the transaction will be garbage collected and further operations will be
rejected.  This optimization can only be applied when there exists a shared fate
between the state machine and the source of proposed values.

\subsection{Capturing Side Effects}

The transaction state machine in Consus records the operations a client wishes
to execute.  These operations drive a state machine that acquires locks,
performs reads, and checks that writes may succeed.  Before a transaction may
commit, these side effects must complete.  The fact that operations are logged
by the state machine does not imply that their side effects have completed.

In order to determine when a transaction has executed and may be passed to the
commit protocol, Consus uses another application of consensus to achieve
agreement among the state machine ensemble.  Consus instantiates a new instance
of the Paxos Synod protocol for each member of the state machine ensemble.
These Synod instances use the ensemble as the set of acceptors.  Each Synod
instance captures the outcome of the transaction for a different member of the
ensemble.  Thus, the execution of a transaction---specifically its side
effects---is captured in these Synod instances.

Once a quorum of these Synod instances capture a successful execution, the
transaction can proceed with the global commit algorithm.  This guarantees that
the entire transactions log is recoverable so long as any quorum of the
transaction ensemble remains live.

This structure is intended to prevent a case where each individual operation in
the transaction log is replicated to a quorum of the ensemble, but no one server
knows the entire log.  Because the entire log is necessary to invoke the commit
protocol, it is necessary to ensure that the log is not only durable to a
minority of failures, but that the complete log is fully replicated on a
majority of the ensemble.  Table~\ref{tab:paxos-pattern} shows a simple example
of a log that is durably replicated, but that does not uphold the invariants
necessary for Consus.

There is a somewhat obscure corner case in this configuration where a minority
of servers become unreachable, but there is no majority that have confirmed
having fully replicated the commit log.  In practice, this is most likely to
happen during unanticipated failures of the system; in theory, it could happen
with an adversarial network.  Accounting for this case makes Consus robust to
programming errors.

\begin{table}[t]
    \centering
    \begin{tabular}{|r|ccccc|}
        \hline
        \T\B        & $S_1$ & $S_2$ & $S_3$ & $S_4$ & $S_5$ \\
        \hline
        \T \code{begin()} & \tickmark & \tickmark & \tickmark & & \\
        \code{read(x)}    & \tickmark & & \tickmark & \tickmark & \\
        \code{write(x)}   & & \tickmark & & \tickmark & \tickmark \\
        \B \code{commit()}& \tickmark & & & \tickmark & \tickmark \\
        \hline
    \end{tabular}
    \caption{An example where every transaction operation is learned by a quorum
    of the servers (3), but no operation is learned by all servers, and no
    server learns of all operations.}
    \label{tab:paxos-pattern}
\end{table}

To overcome this problem case, the instances of the Synod protocol will accept a
value that indicates that a server has failed.  Servers may take up a ballot on
their own or other servers' instances of the Synod protocol and mark that a
server is failed.  Invoking Synod instances in this manner ensures that all
servers will only ever learn a \code{success} or a \code{failure} value for an
instance of the Synod protocol The invariant upheld by Consus is that a server
may propose any value for its own Synod instance, but may only propose
\code{failure} for other servers.  This allows any server to unilaterally decide
that a data center aborts a transaction, but requires a majority of servers
agree that the data center can commit the transaction.

The additional Synod instances also allow the transaction garbage collection
mechanism referred to earlier to safely abort local transactions.  The garbage
collector cannot directly propose to abort a transaction because the client is
the only entity allowed to append operations to the transaction's log.  Instead,
the garbage collector can prevent the transaction from ever making progress by
proposing \code{failure} to each Synod instance and running the protocol until
they complete.  Because this only ever happens on transactions that have
expired, it will not delay their execution.

\subsection{Implicit Phase 1 Paxos}

Consus instantiates multiple instances of Paxos and the Paxos Synod protocol per
transaction.  This is in contrast to traditional uses of Paxos where there is
one long-lived instance of Paxos.  Consequently, Phase 1 of Paxos is invoked
much more often than would otherwise happen.  Because Phase 1 includes durably
logging on remote machines, it can be expensive, and to invoke it multiple times
per transaction adds unnecessary overhead.

Often, one member of a Paxos ensemble will be a suitable default leader for the
Paxos group because it is through its actions that the group is created.  In
such an instance where the entity driving the Paxos group is known in advance,
Consus implicitly sets the entity as having lead a successful Phase 1 ballot.
This avoids the network latency associated with the proposer actually following
Phase 1 of the protocol and the I/O cost of each acceptor durably recording its
Phase 1 promise.  When both are almost surely to succeed, actually executing the
protocol along the regular path is wasteful.  In the rare event that the first
proposer for a Paxos instance is not leading the implicitly chosen ballot, the
proposer could have thrashed with the likely proposer; the implicit Phase 1
delays the implicit proposer from thrashing because it will only learn of the
new ballot when one of its proposals is rejected by an acceptor.

This is purely a performance optimization that largely amounts to changing a
variable's initialization.  All members of the Paxos group must still retain the
code paths necessary to perform Phase 1 of Paxos.

\subsection{Recursive Generalized Paxos}

The commit protocol discussed in Section~\ref{sec:inter} presents each data
center as a singular entity participating in the protocol and running an
instance of Generalized Paxos.  Of course, if this were actually the case, a
single server failure in one data center would make an entire data center appear
offline, introducing more failure handling than would otherwise be desirable.

Consus makes the acceptors for the commit protocol's Generalized Paxos protocol
fault tolerant by sequencing its inputs through a data center-local Generalized
Paxos instance.  Running the global protocol's acceptor on top of a local
replicated state machine immediately makes the whole commit protocol withstand
the failure of a single server without downtime; running it as a generalized
state machine admits more concurrency and availability than would otherwise be
available.

In a standard Paxos-replicated state machine, a single member of the ensemble
operates as the leader for a ballot and proposes values using the authority of
that ballot.  If this member fails, another member of the ensemble must lead a
higher ballot to continue proposing values to the cluster.  This results in high
latency during the transition, and deciding when a server is actually
unavailable rather than executing slightly behind the others is a non-trivial
task to do both quickly and without introducing thrashing between leaders.

In a Generalized Paxos-replicated state machine, any member of the ensemble may
propose any value by directly adding it to its local acceptor's partially
ordered set (poset) of values.  The value becomes accepted when the value
appears in the greatest-upper-bound of the values accepted by a quorum of
acceptors.  Consequently, a value may be chosen once it reaches a quorum of the
ensemble, without funneling the value through any single machine in the
ensemble.

In a recursive Generalized Paxos-replicated state machine, an {\em inner state
machine} runs on top of $N$ ensembles running {\em outer state machines}, each
of which simulates a single acceptor for the inner state machine.  Messages that
would normally be sent to a single acceptor in Generalized Paxos are wrapped in
a proposal and sent to the outer state machine that simulates that acceptor.
Likewise, messages that are normally sent to all acceptors are wrapped in
proposals to all outer state machines.  Each of these outer state machines
learns a partially ordered set of messages that serve as the input to the
acceptor that it represents for the inner state machine.  Any messages generated
by the inner state machine are sent to the requisite outer state machine
ensembles.

The partially ordered set of messages in recursive Generalized Paxos permits a
higher degree of concurrency than a total over messages would permit.  By
default any pair of messages will be ordered in the poset of messages unless a
rule specifically exists that allows the messages to be unordered.  The
rules are:

\begin{itemize}
    \item Phase 1B messages are always unordered with respect to other Phase 1B
        messages for the same ballot.  These messages signal an acceptor
        following a new ballot and the order in which the acceptors follow the
        ballot is not significant.
    \item Phase 2B messages are unordered if there exists a greatest upper bound
        for the poset of the accepted values and the GUB is not $\bot$.  In
        Consus each of these values is a poset of results used in the commit
        protocol.  These messages can commute because updating the acceptor's
        accepted value to the one contained in the Phase 2B message can only
        extend the learned value and cannot violate the safety invariants of
        Generalized Paxos.
    \item Proposals are ordered using the same ordering rules used for the
        relation over the values being proposed.  In Consus, this means that
        proposed results may always commute, while proposed retractions will
        never commute.
    \item Proposals may commute with Phase 2B messages so long as the value
        being proposed is unordered with respect to every element in the
        accepted value contained within the Phase 2B message; again, this
        ordering uses the ordering relation over the inner state machine's
        poset elements.
\end{itemize}

These constraints are sufficient to enable recursive Generalized Paxos to
efficiently decide the commit results without sending the results through any
single machine for sequencing.  The inner state machine is implicitly
initialized to follow a ballot from the origin data center for a transaction.
Then, absent a deadlock-triggered retraction, any number of Phase 2B and
Proposal messages may arrive for the inner state machine and all will be
unordered with respect to each other.  At a high level of abstraction, each
inner state machine is maintaining a set of \code{commit} or \code{abort}
results and retractions for each data center.  Each outer state machine
maintains a copy of this inner state machine.  At any instant in time the outer
state machines may contain a different set of \code{commit} or \code{abort}
results; however, the set will always be a subset what could be learned by a
global observer who can see every message.  A single retraction will force the
outer state machines to converge on a single agreed-upon sequence to the inner
state machine before continuing, thus forcing the inner state machine's values
to converge as well.  It is an open question whether the ordering over the inner
state machine's input could be further relaxed to admit more concurrency; it is
certainly worth investigating more relaxed constraints alongside an
investigation of ways to reduce the likelihood of abort upcalls in the
transaction execution engine.

\section{Implementation}
\label{sec:impl}

The current Consus implementation is approximately \unit{26}{\kilo} lines of
code that depends upon more than \unit{44}{\kilo} lines of supporting code that
was originally written as dependencies of HyperDex.  The Consus and HyperDex
codebases are distinct.  Because HyperDex was written with a different set of
assumptions from Consus, with a different set of desiderata, the HyperDex
implementation was not a useful starting point on which to build Consus.

The biggest difference between HyperDex and Consus is in its approach to fault
tolerance.  HyperDex makes an $f+1$ fault tolerance assumption, where each unit
of $f+1$ nodes can withstand a concurrent failure of any $f$ of those nodes.
The implementation could not stand a concurrent failure of all $f+1$ nodes
without possibly losing data.  Consus assumes that any or all nodes may fail and
resume and its implementation has made this assumption from day one.
Practically, this means that Consus is more conservative in its approach to data
handling, opting to log data to disk rather than rely purely upon replication
for fault tolerance.  If Consus is to fluidly run in multiple data centers in
the presence of failures, it must be able to run and restart in a single data
center without incident, operator involvement, or a recovery procedure.

A more subtle result of the difference in fault tolerance assumptions is that
Consus is engineered to hide latency anomalies to the extent possible.  The
Paxos tricks in Section~\ref{sec:paxos} outline some of the methods used to hide
latency.  The implementation also takes as much system coordination off the
critical path as it can.  In HyperDex a replicated coordinator maintains group
membership for the system.  This coordinator is on the critical path for failure
recovery in HyperDex and must issue a new configuration after each failure to
enable value-dependent chaining to route around the failure.  Consus employs a
similar replicated coordinator, but keeps the configuration out of the critical
path for maintaining availability in the face of failures---Paxos provides
higher availability under failure than HyperDex's chain-based protocol.  The
replicated coordinator is backed by Paxos, so ultimately both Consus and
HyperDex remain available during a failure, but Consus does a better job of
masking a failure and keeping latency consistent during a failure than HyperDex
will.

\subsection{Not Implemented Here}

Consus is an early work-in-progress.  While the commit protocol described in
Section~\ref{sec:inter} and the Paxos optimizations described in
Section~\ref{sec:paxos} are both implemented to the extent described in this
paper, Consus is not yet a fully implemented system.  Specifically, the
isolation described for transaction managers is not implemented, and there is no
concept of schema management.  Applications can write to arbitrary tables with
arbitrary keys and values without restriction.  In a future version of Consus,
the system would provide support for independent schemas, each of which map a
different key-space to the partitions of the key-value store, so that different
skewed workloads can coexist side-by-side.

In a similar vein, Consus has only partially implemented fault tolerance code
paths.  Most modules can be written as a state machine that takes an input and
has side effects.  This state machine can be augmented with a no-input
transition that repeats the side effects necessary to receive an end-to-end
confirmation that the state machine has succeeded in its purpose.  These
transitions are all implemented.  The missing fault tolerance code paths are
largely confined to migrating data when moving partitions from one key-value
store to another.  The key-value stores will not move any of the data and
instead converge to a new mapping almost immediately.  This is simply an
omission from the implementation; it will be added as the system matures.  As
with any moderately-implemented distributed system, more testing is necessary to
ensure the fault tolerance code paths are robust.

Finally, the Consus implementation is not sufficiently mature for a thorough
evaluation.  While it is mature enough to demonstrate the functional basics of
the system, and the core code paths necessary for steady state functioning do
exist, the system needs further development in order to reach a state where a
thorough and proper evaluation can take place.  This is largely a
performance-related concern:  Any sufficiently large system must undergo a
development phase where performance anomalies are systematically removed in
order to provide consistent, reproducible performance results.  Any evaluation
of the current system is as likely to measure implementation anomalies as it is
to measure the performance of the system's contributions, and is unlikely to
report stable results.

\section{Related Work}
\label{sec:relwork}

Consus sits at the cross-section of distributed systems and transaction
management.  The commit protocol in particular makes a concerted effort to
distinguish between the transaction execution---whose core ideas derive from
existing systems---and the commit protocol---which builds heavily on work done
on consensus algorithms.  Existing approaches can largely be classified into the
following categories:

\paragraph{Two and Three Phase Commit}

The classic 2-phase commit~\cite{2pc} algorithm (2PC) enables a transaction to
commit across multiple hosts in an atomic, all-or-nothing fashion.  Using 2PC, a
client can execute the transaction up to the point where it is ready to commit
on every machine.  If the transaction can successfully commit at every host, the
client may then commit the transaction; otherwise it will abort the transaction.

The classic 2PC algorithm suffers from unavailability when the client that is
coordinating the transaction fails, and breaks down entirely if one of the
servers fails concurrently with the client failure.  The 3-phase
commit~\cite{3pc} (3PC) algorithm can reliably commit in the presence of
failures, but assumes a fail-stop model, which is significantly more than the
assumptions of 2PC.

The Consus commit protocol resembles these other commit protocols in that there
is a defined point at which the transaction can, assuming sufficient liveness,
transition to a committed or aborted state.  Consus uses Paxos, which assumes
only crash failures and asynchronous networks; failures cannot be reliably
detected, and do not need to be reliably detected.

\paragraph{Spanner and Replicated Commit}

Spanner~\cite{spanner} implements a classic 2-phase commit protocol where all
participants to the protocol are made fault tolerant using Multi-Paxos.  The key
contribution of Spanner is a technique called TrueTime that bounds clock
uncertainty to allow fast lockfree reads all around the globe.  Replicated
Commit~\cite{replicatedcommit} flips the layering of 2PC and Paxos such that
each data center runs its own instance of 2PC and uses Paxos to reach consensus
across the 2PC groups.  By switching the layering of 2PC and Paxos, Replicated
commit is able to alter the number of cross-data center round trips necessary to
execute a transaction.  Whether Spanner or Replicated Commit yield a lower
latency is largely a result of the number of data centers and the number of
reads within a transaction.  Consus takes a different approach where a
transaction optimistically executes entirely within one data center, but the
commit protocol will incur exactly three one way communication delays in the
common case.  For transactions that are unlikely to contend in the wide area,
this is a strict improvement on both Spanner and Replicated Commit.

\paragraph{Paxos and Consensus}

Consensus forms the foundation of Consus.  A common consensus algorithm, and the
one used in Consus is Paxos~\cite{paxos}.  Most of the applications of consensus
within Consus used Paxos as the basis of state-machine
replication~\cite{fredrsm}.  Section~\ref{sec:paxos} detailed some optimizations
used within Consus in regards to Paxos.  To our knowledge, the recursive
Generalized Paxos optimization described in that section is the first such
algorithm that uses Paxos to make a Paxos state machine's acceptors more fault
tolerant than a single machine.

Recent work has presented Egalitarian Paxos~\cite{epaxos}, which provides high
performance Paxos in the wide area.  The authors show that Egalitarian Paxos can
provide higher performance in the wide area than other Paxos protocols,
including Generalized Paxos.  Consus' use of Generalized Paxos is much more
specialized than the general replication that Egalitarian Paxos readily
supports in order to tightly control latency.

\section{Discussion}
\label{sec:discuss}

This paper presents a preliminary, and as of yet, untested, description of the
contributions of Consus.  The paper accompanies an open source release of the
complete Consus code base.  This is, in and of itself, an experiment to see to
what extent a system can undergo review and feedback prior to formal
publication.

Often the evaluation of a system is seen as  the single biggest determining
factor in whether a particular piece of work is worthy of publication, or one of
the many papers rejected from a conference.  While the evaluation of a system is
important, it is inherently biased; systems researchers may unconsciously
illuminate the most flattering aspects of a system while overlooking critical
flaws.  Good systems builders will often know what to evaluate in a system
before the system itself is capable of such evaluation.  During construction of
an artifact, they will naturally tend to favor development in these areas and
pay less attention to other aspects of the artifact's development.  These other
areas may be orthogonal concerns, relevant concerns that can be hand-waved away
with black box applications of existing work, or relevant concerns that go
unaddressed in the final evaluation.  Even the most well-intentioned of
researchers have blind spots and the peer review process is intended to
compensate for this by providing objective evaluation of the work.  The extent
to which a system is implemented is often implicitly stated by the paper; more
importantly, the extent to which a system is {\em not} implemented is almost
never stated.  Thus, the peer review process operates through a layer of paper
indirection where there is a description of an artifact and its evaluation,
without any strong insight into the state of the artifact itself or how
accurately the paper reflects the measured artifact.  If the actual artifact is
released at all, it often comes after publication, and possibly through
restricted and less open means than the publication process itself.

The ACM recognizes the importance of artifact evaluation in the publication
process~\cite{arteval}; this pre-publication is an effort to go one step
further.  By opening up Consus to review and feedback from academics and curious
engineers alike, we are able to better understand its strengths, limitations,
and shortcomings prior to any formal publication.  This is almost certainly an
endeavor that is not without some degree of risk to the authors.  Every
publication has the possibility of being ``scooped'' when another publication is
able to make enough of a substantially similar contribution as to render the
original publication untenable.  For authors, there is a real tension between
withholding information in order to ensure that a publication is not scooped,
and publishing information in order to receive feedback on it from ones peers.
This process typically takes place behind closed doors, and often via a layer of
paper indirection.

With Consus, the authors have chosen to release this preliminary description of
the contributions of the system alongside the complete artifacts that will be
evaluated in the eventual peer-reviewed publication.  At the time of this
publication, the artifact itself is not sufficiently robust to provide a
rigorous evaluation platform.  The development and testing necessary for a
thorough evaluation test the limits of a small development team; to delay the
process of outside review until such time that the system is robust for
evaluation could quite literally consume years of developer resources.  Any
review that questions fundamentals of the system could require duplicating much
of that effort.  Releasing the code and a description of the system outside
normal peer review channels can accelerate the critical feedback loop that makes
peer review such a useful tool.  It is our hypothesis that doing so will ease
the eventual formal peer review and publication process; it is our hope that
doing so will serve as a case study in ways to improve the review process as a
whole as artifact evaluation becomes more pertinent to our field.

\section{Conclusion}
\label{sec:concl}

This paper introduces Consus, a new strictly serializable, geo-replicated
key-value store.  The key contribution of Consus is a commit protocol that can
enable commit in three wide area message delays in the common case.  Through the
application of some Paxos optimizations, Consus is able to provide theoretically
better performance than other geo-replicated systems.

This paper lays the ground work for understanding the properties of Consus, and
points to the eventual evaluation of the system.  The Consus source code is
available in parallel to this work in order to facilitate more effective review
of the work, and to permit a broader community to inspect and work with the
artifact prior to a peer-reviewed publication.

\hbadness=10000
\bibliographystyle{plain}
\bibliography{consus}

\end{document}